\documentclass[manuscript]{aastex}
\begin{document}


\title{Orbital and Spin Parameter Variations of Partial Eclipsing Low Mass X-ray Binary X 1822-371}

\author{Yi Chou\altaffilmark{1}, Hung-En Hsieh\altaffilmark{1}, Chin-Ping Hu\altaffilmark{1,2}, Ting-Chang Yang\altaffilmark{1} and Yi-Hao Su\altaffilmark{1}}

\altaffiltext{1}{Graduate Institute of Astronomy, National Central University, Jhongli 32001, Taiwan} 
\altaffiltext{2}{Department of Physics, University of Hong Kong, Pokfulam Road, Hong Kong, China}
\email{Chou: yichou@astro.ncu.edu.tw, Hsieh: m1029003@astro.ncu.edu.tw}

\begin{abstract}
We report our measurements for orbital and spin parameters of X 1822-371 using its X-ray partial eclipsing profile and pulsar timing from data collected by the {\it Rossi X-ray Timing Explorer (RXTE)}. Four more X-ray eclipse times obtained by the {\it RXTE} 2011 observations were combined with historical records to trace evolution of orbital period. We found that a cubic ephemeris likely better describes evolution of the X-ray eclipse times during a time span of about 34 years with a marginal second order derivative of $\ddot{P}_{orb}=(-1.05 \pm 0.59) \times 10^{-19}$ s$^{-1}$. Using the pulse arrival time delay technique, the orbital and spin parameters were obtained from {\it RXTE} observations from 1998 to 2011. The detected pulse periods show that the neutron star in X 1822-371 is continuously spun-up with a rate of $\dot{P}_{s}=(-2.6288 \pm 0.0095)  \times 10^{-12}$ s s$^{-1}$. Evolution of the epoch of the mean longitude $l=\pi /2$ (i.e. $T_{\pi / 2}$) gives an orbital period derivative value consistent with that obtained from the quadratic ephemeris evaluated by the X-ray eclipse but the detected $T_{\pi / 2}$ values are significantly and systematically earlier than the corresponding expected X-ray eclipse times by $90 \pm 11$ s. This deviation is probably caused by asymmetric X-ray emissions.  We also attempted to constrain the mass and radius of the neutron star using the spin period change rate and concluded that the intrinsic luminosity of X 1822-371 is likely more than  $10^{38}$ ergs s$^{-1}$.   
\end{abstract}


\section{Introduction} \label{intro}

X 1822-371 is a typical partial eclipsing low mass X-ray binary (LMXB) with a high inclination angle of $i=82{\degr}.5 \pm 1{\degr}.5$ \citep{hei01}. Its low X-ray to optical luminosity ratio $L_x / L_{opt} \sim 20$ \citep{gri78} in comparison with that a typical LMXB of $\sim 500$ \citep{van95}, and partial eclipse imply that there is an accretion disk corona (ADC; \citealt{whi82}) around the center of the compact object and accretion disk. Its isotropic X-ray luminosity is $\sim 10^{36}$ ergs s$^{-1}$ for an assumed distance of 2.5 kpc \citep{mas82b}, but the intrinsic X-ray luminosity is probably as high as $ \sim 10^{38}$ ergs s$^{-1}$ or even likely close to its Eddington limit \citep{whi82,bur10,bay10}, because of the obscuration. Because of the high inclination angle, the observable X-rays are scattered from the ADC and the X-ray emission region is extended. The radius of the ADC is about half of the accretion disk radius ($2-3 \times 10^{10}$ cm; \citealt{whi82,hel89}). Such a large ADC results that only a part of the X-ray emission region is blocked by the companion during an eclipse.      

Its orbital modulation period of 5.57 h can be observed in the X-ray \citep{whi81}, ultraviolet \citep{mas82b}, optical \citep{sei79} and infrared \citep{mas82a} bands. The orbital variation in the X-ray band consists of a partial eclipse profile plus a smooth broad feature with a minimum about 0.15 cycle prior to the eclipse time \citep{whi81}. This smooth modulation is believed to be caused by obscuration of the ADC by a thick accretion disk rim \citep{whi82}. Using X-ray eclipse times, i.e., taking the minimum intensity in a partial ecliepse profile as the fiducial point, the orbital period and its evolution of X 1822-371 has been monitored since the early 80's.  \citet{hel90} first reported a significant orbital period derivative of $\dot{P}_{orb}=(2.19 \pm 0.58) \times 10^{-10}$ s s$^{-1}$ from X-ray eclipses. The quadratic ephemeris of X 1822-371 was further updated using more X-ray observations \citep{hel94,par00,bur10,iar11}, and the latest reported orbital period derivative is $\dot{P}_{orb}=(1.514 \pm 0.080) \times 10^{-10}$ s s$^{-1}$ \citep{iar11}. Moreover, \citet{sei79} discovered a 5.57 h orbital modulation in the optical counterpart of X 1822-371. The optical light curve shows a broad partial eclipse profile, indicating that the system has a large accretion disk with a radius of  $r_d /a = 0.58$ \citep{hel89} where $r_d$ is the disk radius and $a$ is the binary separation. Using the minimum intensity of the light curve as the fiducial point, the optical/UV ephemeris has kept updating since \citet{cha80} reported the first linear ephemeris. The orbital period derivative derived from the eclipse times of optical/UV light curves was first detected by \citet{bap02} and the quadratic ephemeris was further refined by \citet{bay10} and \citet{iar11}. Basically, the quadratic ephemerides derived from the X-ray and optical/UV bands are consistent with each other, except that the optical eclipse times systematically lag the X-ray eclipse times by about 2-3 min. This phenomenon was first noted by \citet{whi81} and then further confirmed and more precisely evaluated in later observations \citep{hel89,hel90,iar11}. The time lag is likely caused by the different regions of optical and X-ray emissions in the system. The X-rays are from the ADC whereas the optical emissions are from an asymmetic accretion disk \citep{hel89}. However, the orbital period derivatives derived from both X-ray and optical/UV observations are more than 1000 times larger than that evaluated using the mass transfer driven by gravitational radiation and magnetic braking with the assumption of total mass conservation in the binary system \citep{bur10}. It is possible that more than 70\% of mass loss from the companion is expelled from the system \citep{bur10}. In contrast, it also could be caused by short-term departures from long-term evolutionary trends, such as the magnetic cycles of the companion changing its quadrupole moment and resulting in the orbital period variation \citep{hel90}.   

The accretor of X 1822-371 was identified to be a neutron star when \citet{jon01} discovered its 0.59 s pulsation from a {\it Rossi X-ray Timing Explorer (RXTE)} observation made in 1998. Some of the orbital parameters were revealed by pulsar timing, including a projected semimajor axis $a_x \sin i=$1.006(5) lt-s for the neutron star and a circular orbit with eccentricity less than 0.03. Combined with the pulsation period detected from another {\it RXTE} observation in 1996, they found a spin-up rate of $(-2.85 \pm 0.04) \times 10^{-12}$ s s$^{-1}$. The magnetic field derived from the X-ray spectrum is $\sim (1-5) \times 10^{12}$ G, implying an intrinsic X-ray luminosity of $\sim (2-4) \times 10^{37}$ erg s$^{-1}$ \citep{jon01}. The spin period change rate was further traced by \citet{jai10,sas14}, and \citet{iar15}. In addition, \citet{sas14} and \citet{iar15} both claimed a discovery of cyclotron resonant scattering features but with very different center energies. \citet{sas14} detected a significant cyclotron resonant scattering feature at $33 \pm 2$ keV from a {\it Suzaku} observation in 2006, implying that the magnetic field of the neutron star is $(2.8 \pm 0.2) \times 10^{12}$ G and the luminosity is $\sim 3 \times 10^{37}$ erg s$^{-1}$. However, \citet{iar15} observed a cyclotron resonant scattering feature at 0.7 keV from an {\it XMM-Newton} observation in 2001, and no similar feature can be seen around 33 keV from all available {\it INTEGRAL} observations. This 0.7 keV resonant feature indicates that the magnetic field on the neutron star is only $(8.8 \pm 0.3) \times 10^{10}$ G and the intrinsic luminosity is as high as $\sim 10^{38}$ erg s$^{-1}$, close to its Eddington limit. \citet{iar15} also argued that the magnetic field proposed by \citet{sas14} is too high to allow accretion because the magnetospheric radius would be larger than the corotation radius.   

In this paper, we present our analysis results for measuring the orbital and spin parameters of X 1822-371 using the archival {\it RXTE} data (section~\ref{dar}). Combined with historical records, a cubic ephemeris was established for better describing the X-ray eclipse times of X 1822-371 for a time span of $\sim 34$ years (section~\ref{ea}). In addition to the spin period, the orbital parameters, including orbital period, projected semimajor axis ($a_x \sin i$), and epochs of mean longitude $l=\pi/2$ ($T_{\pi / 2}$) were evaluated for the individual {\it RXTE} observations from 1998 to 2011 using pulsar timing (section~\ref{pa}). The detected $T_{\pi / 2}$ values are systematically earlier than the corresponding X-ray eclipse times by $\sim 90$ s. The analysis results, including, orbital period evolution, the possible implications of the difference between X-ray eclipse time and  $T_{\pi / 2}$, and using the detected spin-up rate to constrain the mass and radius of neutron star, are discussed in section~\ref{dis}.

\section{Observations} \label{obs}

The {\it RXTE} observations for X 1822-371 were made in 1996 (Obs ID:10115), 1998 (Obs ID:30060), 2001 (Obs ID:50048 and 60042), 2002-2003 (Obs ID: 70037), and 2011 (Obs ID: 96344 and 96377). The data used in this study were collected by {\it RXTE} Proportional Counter Array \citep[PCA;][]{jah96}, consisting of five proportional counter units (PCUs) with a total photon collecting area of 6500 cm$^2$ sensitive to  photons in the energy range of 2-60 keV. The X-ray light curves for measuring the X-ray eclipse times were directly obtained from the {\it RXTE} standard products, collected in Standard 2 mode with a time resolution of 16 s. The PCA data used for analyzing pulsar timing were recorded either in the GoodXenon mode with a time resolution of 1 $\mu$s or in the Generic Event mode with a time resolutions of 16 $\mu$s or 125 $\mu$s.

\section{Data Analysis and Results} \label{dar}

\subsection{Orbital Ephemeris from X-ray Eclipse Times} \label{ea}
\begin{figure}
\plotone{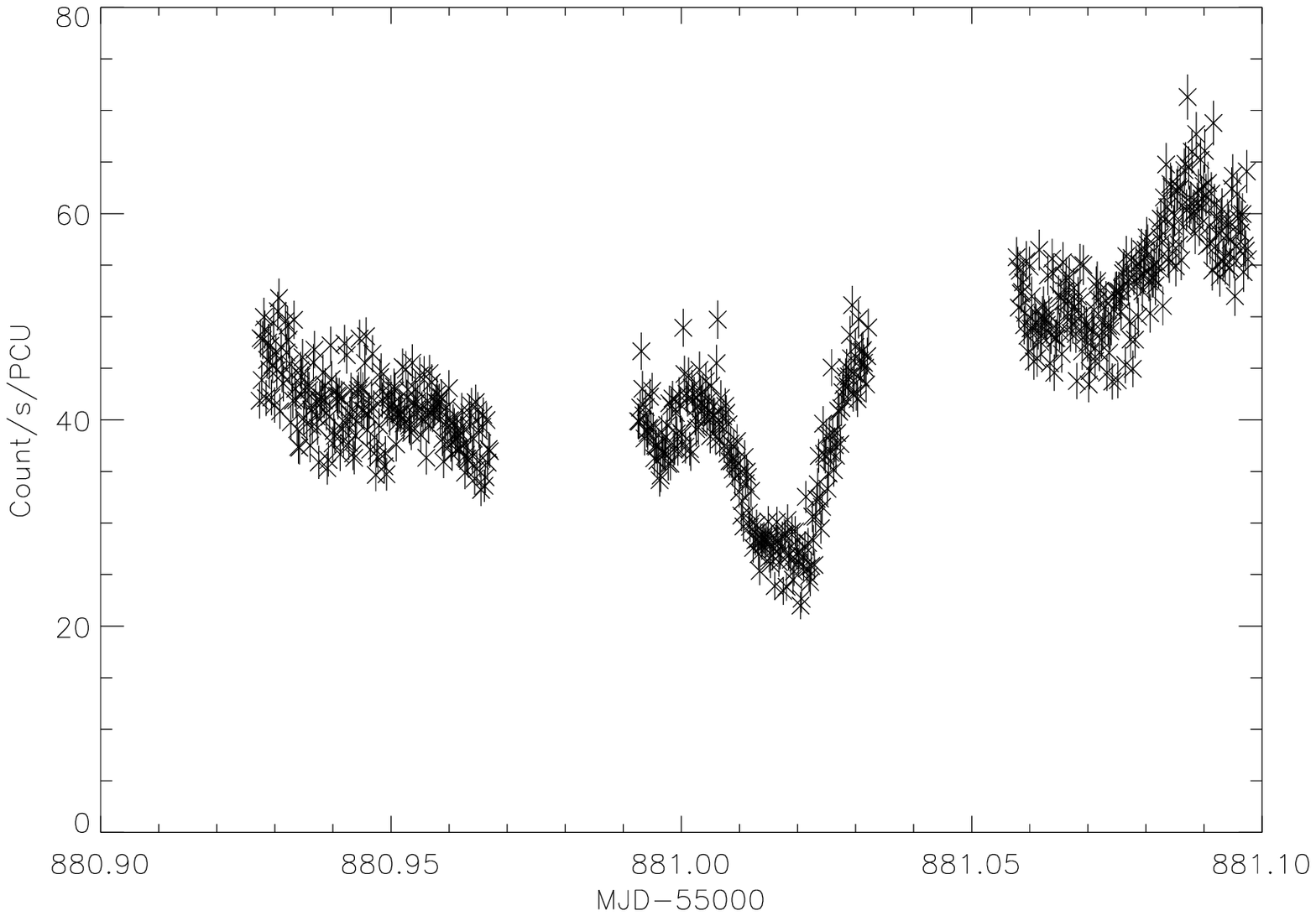}
\caption{A typical 2-9 keV light curve with eclipse profile detected by the RXTE/PCA observed on November 15, 2011. \label{lc}}
\end{figure}

A complete journal of X-ray eclipse times for X 1822-371 prior to this work has been listed in Table 2 of \citet{iar11}. In this paper, we further added more X-ray eclipse times detected by {\it RXTE} 2011 observations to improve the orbital ephemeris. The X-ray light curves were directly retrieved from the standard data products (StdProds) of archival {\it RXTE} data. The 2-9 keV, background-subtracted light curves collected by the PCA with a time resolution of 16 s were selected for analysis. The time column ``BARYTIME'', whose values have been corrected to the barycenter of the solar system, was used in the light curves. There are four eclipse minima, i.e., the fiducial points that allow us to determine X-ray eclipse times, can be clearly seen in the whole {\it RXTE} 2011 observations. A typical 2-9 keV light curve with eclipse profile detected by the {\it RXTE}/PCA is shown in Figure~\ref{lc}.

To extract X-ray eclipse times and the corresponding uncertainties, we adopted a method analogous to that used in \citet{par00}. The light curve around the eclipse profile was fitted with three models, i.e., a Gaussian plus a constant, a linear function, and a quadratic function. The X-ray eclipse time was determined by averaging the Gaussian centroid values from these three models and the uncertainty was evaluated, similar to that used in \citet{bur10}, the half of maximum range span of these three centroid values. The X-ray eclipse times detected by {\it RXTE} 2011 observations are listed in Table~\ref{etime}.    

\begin{figure}
\plotone{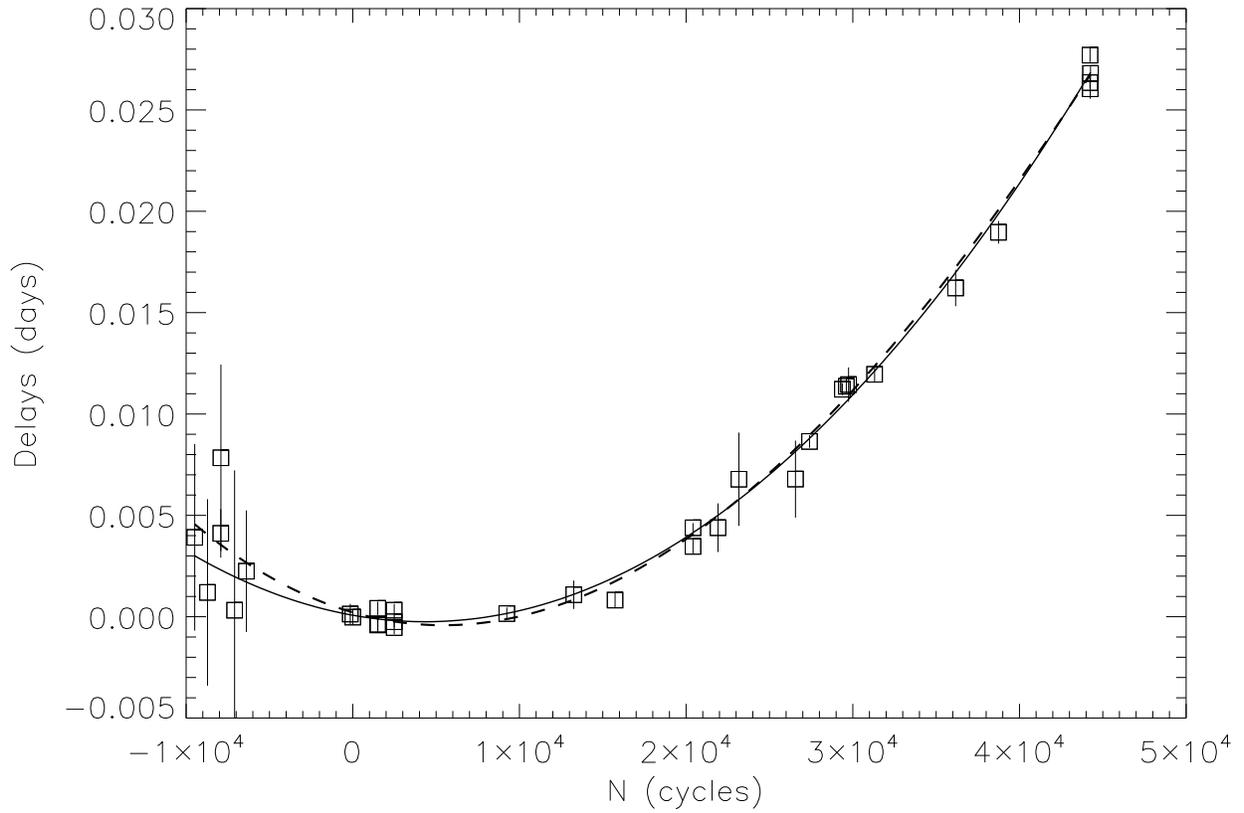}
\caption{Evolution of X-ray eclipse time delay relative to the linear ephemeris proposed by \citet{hel94} and the corresponding best fits of the quadratic model (solid line) and cubic model (dashed line). \label{o-c}}
\end{figure}

Combined with historical records of eclipse times listed in Table 2 of \citet{iar11}, we adopted the observed-minus-calculated (O-C) method to fit the time delays between the observed X-ray eclipse times and a linear ephemeris proposed by \citet{hel94} with a quadratic function of cycle counts under the assumption that the orbital period derivative is a constant during the whole $\sim$34 years' of time span, from 1977 to 2011. The evolution of time delays and the best-fitted quadratic curve are shown in Figure~\ref{o-c}, and the corresponding parameters are listed in Table~\ref{efit}.  Because the reduced $\chi^2$ ($\chi^2_{\nu}$) value is significantly larger than 1, to conservatively estimate the uncertainties, all errors of the parameters were scaled by a factor of $\sqrt{\chi^2_{\nu}}$. We therefore obtained a period derivative of $\dot{P}_{orb}=(1.464 \pm 0.041) \times 10^{-10}$ s s$^{-1}$, which is consistent with but a little smaller than the value reported by \citet{iar11}, and the updated quadratic ephemeris is

\begin{equation}\label{e_eph2}
T_N=45614.80949(17)MJD/TDB+0.232108983(91)N+(1.700 \pm 0.048)\times10^{-11}N^2,
\end{equation}
\noindent where $N$ is the cycle count number.

 However, from the reported orbital period derivatives of X 1822-371 evaluated using the X-ray eclipses listed in Table~\ref{ep_dot}, we found that the period derivatives decrease with increasing observed time span, even though these values are consistent with each other. This implies that there is a negative second order orbital period derivative $\ddot{P}_{orb}$ of about $-1.4 \times 10^{-19}$ s$^{-1}$, evaluated from the reported $\dot{P}_{orb}$.\footnote{We assumed that $\ddot{P}_{orb}$ is a constant and considered the reported $\dot{P}_{orb}$ values In Table~\ref{ep_dot} as the mean orbital period derivatives of the corresponding time spans. The mean orbital period derivative can be expressed as $\overline{\dot{P}_{orb}}(t)=\dot{P}_{0,orb}+\onehalf \ddot{P}_{orb} \times t$ where $t$ is the time span.} We therefore fitted the time delays with a cubic function of cycle counts. The best-fitted cubic curve is shown in Figure~\ref{o-c} and the parameters are listed in Table~\ref{efit}. Comparing the quadratic and cubic models, the F-test gave an F-value of 3.18, indicating the cubic model improves the fitting with a confidence level of 91.47\%. Similar to the quadratic fitting, we multiplied all the errors from the fitting with a factor of $\sqrt{\chi^2_{\nu}}$ and obtained a marginal second order orbital derivative of $\ddot{P}_{orb}=(-1.05 \pm 0.59) \times 10^{-19}$ s$^{-1}$, consistent with that estimated from reported $\dot{P}_{orb}$ values. Thus, we established a cubic ephemeris that likely better describes the X-ray eclipse time evolution in this 34-year time span as

\begin{equation}\label{e_eph3}
T_N=45614.80964(18)MJD/TDB+0.232108780(54)N+(2.25 \pm 0.31)\times10^{-11}N^2-(8.2 \pm 4.6) \times 10^{-17}N^3.
\end{equation}

\subsection{Orbital and Spin Parameters from Pulsar Timing} \label{pa}

Besides X-ray eclipse times, orbital and spin parameters can also be precisely determined using the pulse arrival time delay technique. The PCA science event files described in section~\ref{obs} were adopted in the following analysis. All X-ray photon arrival times were first corrected to the barycenter of solar system. Following \citet{jon01}, we selected the events within the energy range of 9.4-22.7 keV and divided them into data segments of $\sim$1500 s. We derived the power spectra of all data segments using the $Z^2_1$ test \citep{buc83} and chose only those with significant pulsation detection (more than 95\% confidence level) for further analysis.  Because of orbital motion, the orbital Doppler effect can be clear seen in variations of the detected spin frequencies from the power spectra. Furthermore, because time span of the whole data set is about 15 years, significant linear frequency drift caused by spin frequency derivatives can be also observed. By the unweighted fitting with the circular orbit model plus a linear function for the detected spin frequencies, we obtained preliminary orbital and spin parameters including orbital period $P_{orb}=20054.34601133(78)$ s, projected semimajor axis $a_{x} \sin i=1.003 \pm 0.033$ lt-s, and spin period derivative $\dot{P}_{s}=(-2.598 \pm 0.031) \times 10^{-12} $ s s$^{-1}$, which are close to the values reported by \citet{jon01}, \citet{jai10}, \citet{sas14} and \citet{iar15}.

\begin{figure}
\plotone{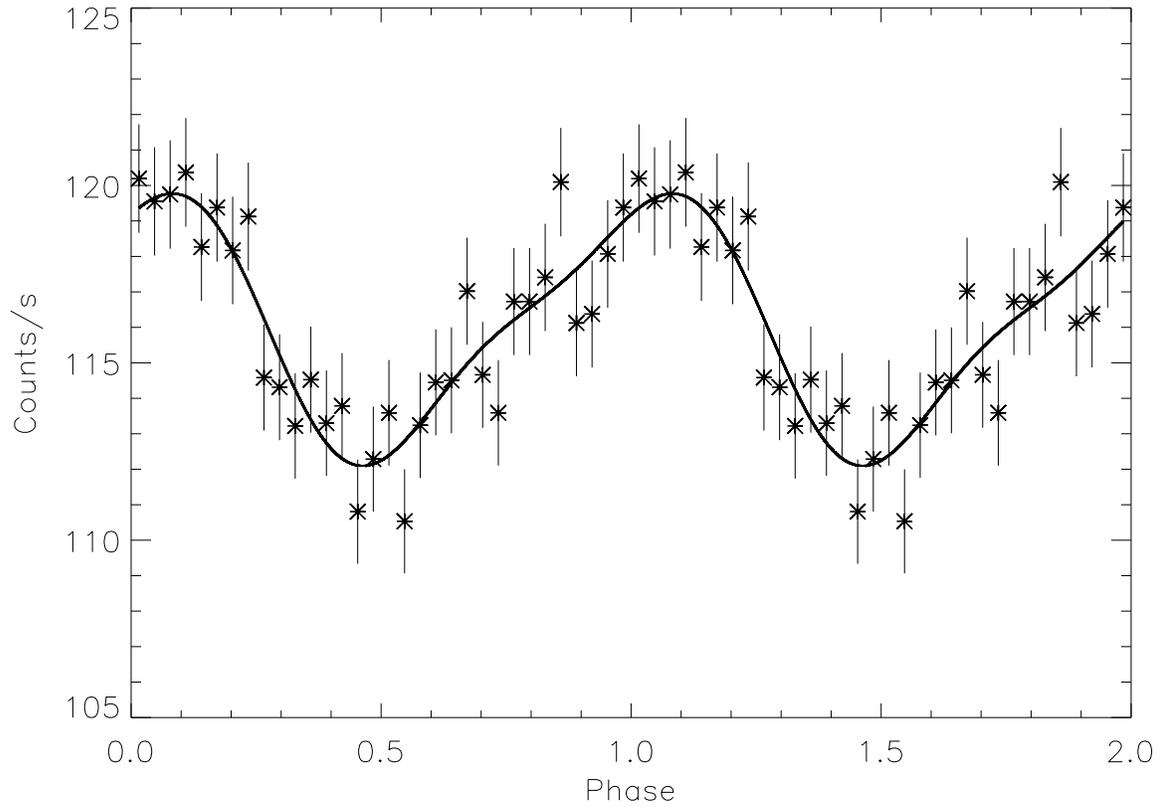}
\caption{Typical pulse profile of a data segment observed on July 4, 2001. The solid curve is the multiple sinusoidal fitting result of the pulse profile. \label{pulse_profile}}
\end{figure}

More precise orbital and spin parameters can be obtained using the pulse arrival time delay technique.  The event arrival times $t_i$ of each data segment were folded with a time-variable frequency caused by the orbital Doppler effect of the circular orbit to obtain the event phase $\phi_i$ as

\begin{eqnarray}\label{fold}
\phi_i & = &frac  \int_{T_0}^{t_i}\left\{\nu_0+\nu_0{{2\pi a_x \sin i} \over {c}}f_{orb}\sin \left[{{2 \pi f_{orb}(t-T_{\pi / 2})} }\right]\right\} dt  \\
\nonumber & = & frac \left\{\nu_0(t_i-T_0)-\nu_0{{a_x \sin i} \over {c}}\cos \left[{{2 \pi f_{orb}(t_i-T_{\pi / 2})}}\right]+\nu_0{{a_x \sin i} \over {c}}\cos \left[{{2 \pi f_{orb}(T_0-T_{\pi / 2}) }}\right]\right\},
\end{eqnarray}

\noindent where $f_{orb}= 1/P_{orb}$, $a_x \sin i$, $T_{\pi /2}$, $\nu_{0}$ and $T_{o}$ are the orbital frequency of the binary system, projected semimajor axis, epoch of 90$^\circ$ mean longitude, spin frequency of the neutron star, and epoch of phase zero of pulsation, respectively. It is equivalent to correcting the event times to the barycenter of the binary system and then folding with a constant frequency. The pulse profile of each data segment was made by binning the event phases. Figure~\ref{pulse_profile} shows the typical pulse profile of a data segment. The profile was fitted with a multiple sinusoidal function, by keeping adding higher order harmonic term to the model until the F-test indicates that adding higher order harmonics has no significant improvement on the fitting (with confidence level less than 90\%).  The peak of the best-fitted profile was selected as the fiducial point in the following analysis. The uncertainty of the pulse phase was evaluated by $10^4$ runs of a Monte Carlo simulation. 

If the orbital and spin parameters in Eq~\ref{fold} were exactly the true orbital and spin parameters of the binary system, the pulse phases should have been aligned on phase zero. In contrast, if the guess parameters slightly deviate from the true ones, the pulse phase drift can be expressed (in first-order approximation) as 

\begin{eqnarray}\label{orb_corr}
\delta \phi(t) &=&  \bigl \{  -(t-T_0^{(0)})+A^{(0)} \cos \bigl [ {{2\pi f_{orb}^{(0)} (t-T_{\pi/2}^{(0)})} } \bigr]  - A^{(0)} \cos \bigl [ {{2\pi f_{orb}^{(0)} (T_0^{(0)}-T_{\pi/2}^{(0)})} } \bigr]\bigr \} \delta \nu_0 \\
\nonumber & & + \bigl \{ \nu_0^{(0)}+2\pi A^{(0)} \nu_0^{(0)}f_{orb}^{(0)}\sin \bigl [ {{2\pi f_{orb}^{(0)} (T_0^{(0)}-T_{\pi/2}^{(0)})} } \bigr]     \bigr\} \delta T_0\\
\nonumber & & +\bigl \{ \nu_0^{(0)}\cos \bigl [ {{2\pi f_{orb}^{(0)} (t-T_{\pi/2}^{(0)})} } \bigr] - \nu_0^{(0)} \cos \bigl [ {{2\pi f_{orb}^{(0)} (T_0^{(0)}-T_{\pi/2}^{(0)})} } \bigr] \bigr \} \delta A \\
\nonumber & & + \bigl \{-2\pi\nu_0^{(0)}A^{(0)}(t-T_{\pi/2}^{(0)}) \sin  \bigl [ {{2\pi f_{orb}^{(0)} (t-T_{\pi/2}^{(0)})} } \bigr] \\
\nonumber & & + 2\pi\nu_0^{(0)}A^{(0)}(T_0^{(0)}-T_{\pi/2}^{(0)}) \sin  \bigl [ {{2\pi f_{orb}^{(0)} (T_0^{(0)}-T_{\pi/2}^{(0)})} } \bigr]\bigr \} \delta f_{orb}\\
\nonumber & & + \bigl \{ 2\pi \nu_0^{(0)}A^{(0)} f_{orb}^{(0)}  \sin  \bigl [ {{2\pi f_{orb}^{(0)} (t-T_{\pi/2}^{(0)})} } \bigr] - 2\pi \nu_0^{(0)}A^{(0)} f_{orb}^{(0)}  \sin  \bigl [ {{2\pi f_{orb}^{(0)} (T_0^{(0)}-T_{\pi/2}^{(0)})} } \bigr]\bigr \} \delta T_{\pi/2},
\end{eqnarray}

\noindent where $A \equiv a_x \sin i / c$, the parameters with a superscripts (0) represent the guess parameters, and the parameters with $\delta$ are the differences between true and guess parameters. The parameter corrections can be obtained fitting fitting the phase drift with time using Eq~\ref{orb_corr}. This correction process can be iterated until the corrected values are much smaller than the corresponding errors.

To trace evolution of the orbital and spin parameters, we divided the selected data segments into 9 data sets according to their observation times with no time span of each data set longer than $\sim$15 d and no less than 7 data segments for each data set to resolve the spin and orbital parameters. Data segments not satisfying the conditions were excluded from constraining the orbital and spin parameters because of insufficient degrees of freedom. Different initial guess parameters were applied to the data sets to fold the event times using Eq~\ref{fold} and made the pulse profiles for the data segments. The initial guess values of the spin frequency were evaluated using the ephemeris proposed by \citet{jai10}; the projected semimajor axis obtained by \citet{jon01} was adopted as the initial guess value for all the data sets, and the initial guess values for $T_{\pi/2}$ was calculated using the cubic ephemeris obtained from partial eclipses (Eq~\ref{e_eph3}). The ephemeris obtained from partial eclipses can give a very precise estimation of the orbital period because of the long time span; therefore, the orbital frequencies were evaluated using the cubic ephemeris and kept as constants for parameter corrections (i.e. $\delta f_{orb}$=0 in Eq~\ref{orb_corr}).

The parameter correction process described above was applied to each data set.  Table~\ref{para} listed the best-fit parameters for individual data set. To improve the fitting significantly, an additional spin frequency derivative term (i.e. $1/2 \dot \nu_s (t_i - T_0)^2$) was required by some data sets, evaluated by F-test with confidence levels larger than 95\%. We also added a small eccentricity to our orbital model and found no significant eccentricity can be detected with a 2$\sigma$ upper limit of 0.04.

\begin{figure}
\plotone{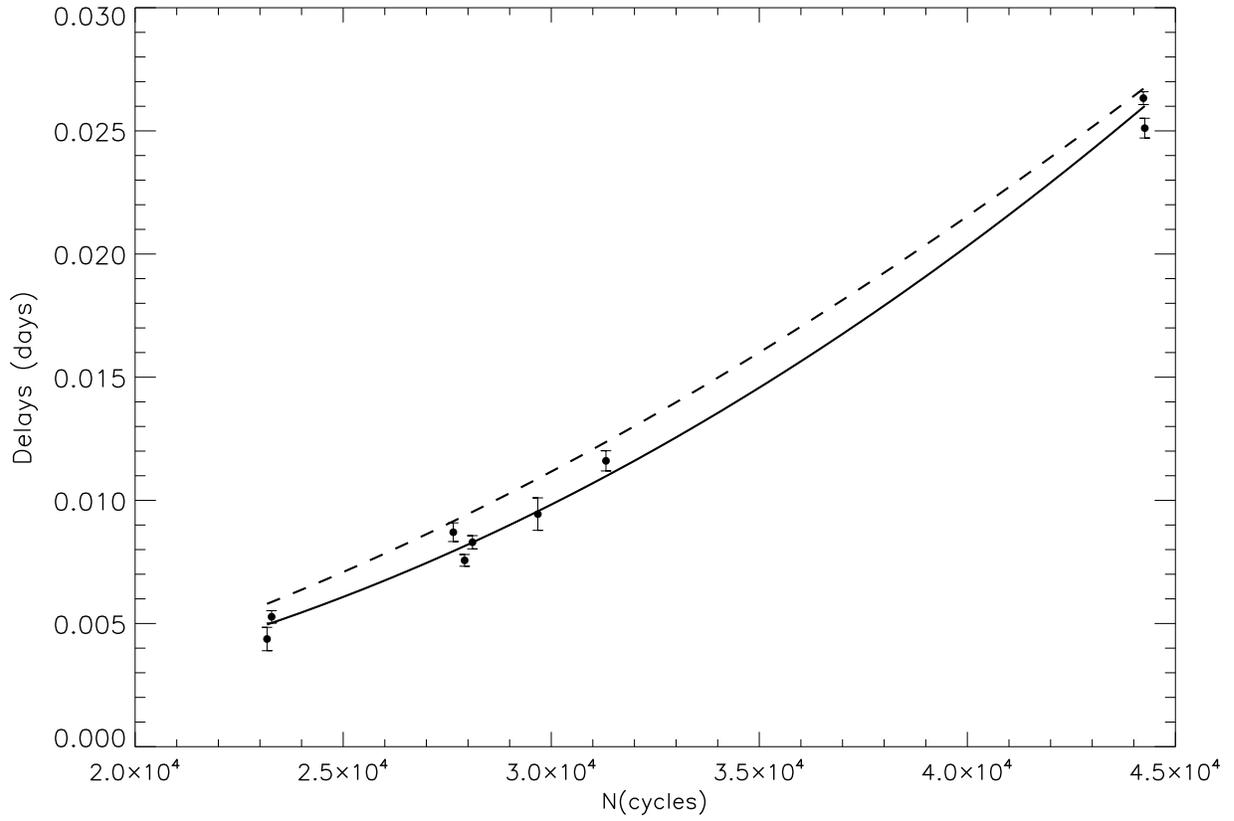}
\caption{Evolution of time delay of $T_{\pi /2}$ relative to the linear ephemeris proposed by \citet{hel94} and the corresponding best fits of the quadratic model (solid line). For comparison, the dashed line represents the best cubic model of X-ray eclipse times. \label{to-c}}
\end{figure}

As can see from the projected semimajor axis values listed in Table~\ref{para}, there is no significant change from 1998 to 2011 with a weight average value of 1.0021(30) lt-s, consistent with the value proposed by \citet{jon01}. Moreover, the measured $T_{\pi /2}$ values can provide an independent approach besides eclipse to probe the orbital period evolution. We therefore applied the similar O-C method used for X-ray eclipse times to the $T_{\pi /2}$ values listed in Table~\ref{para}, and found that a quadratic function can well describe the evolution of the $T_{\pi /2}$ values with orbital derivative of $(1.72 \pm 0.45)\times 10^{-10}$ s s$^{-1}$ (Figure~\ref{to-c}), consistent with the value obtained from the quadratic ephemeris of eclipses. No higher order orbital derivatives can be detected probably because of relatively small time span compared with the eclipses. Therefore, we establish the quadratic ephemeris for $T_{\pi /2}$ as    

\begin{equation}\label{t_eph2}
T_N=45614.8117(60)MJD/TDB+0.2321089(16)N+(2.00 \pm 0.53)\times10^{-11}N^2.
\end{equation}

\noindent All errors of the parameters have been scaled by a factor of $\sqrt{\chi^2_\nu}$ as we did for establishing the ephemerids of X-ray eclipse times in section~\ref{ea}.

\begin{figure}
\plotone{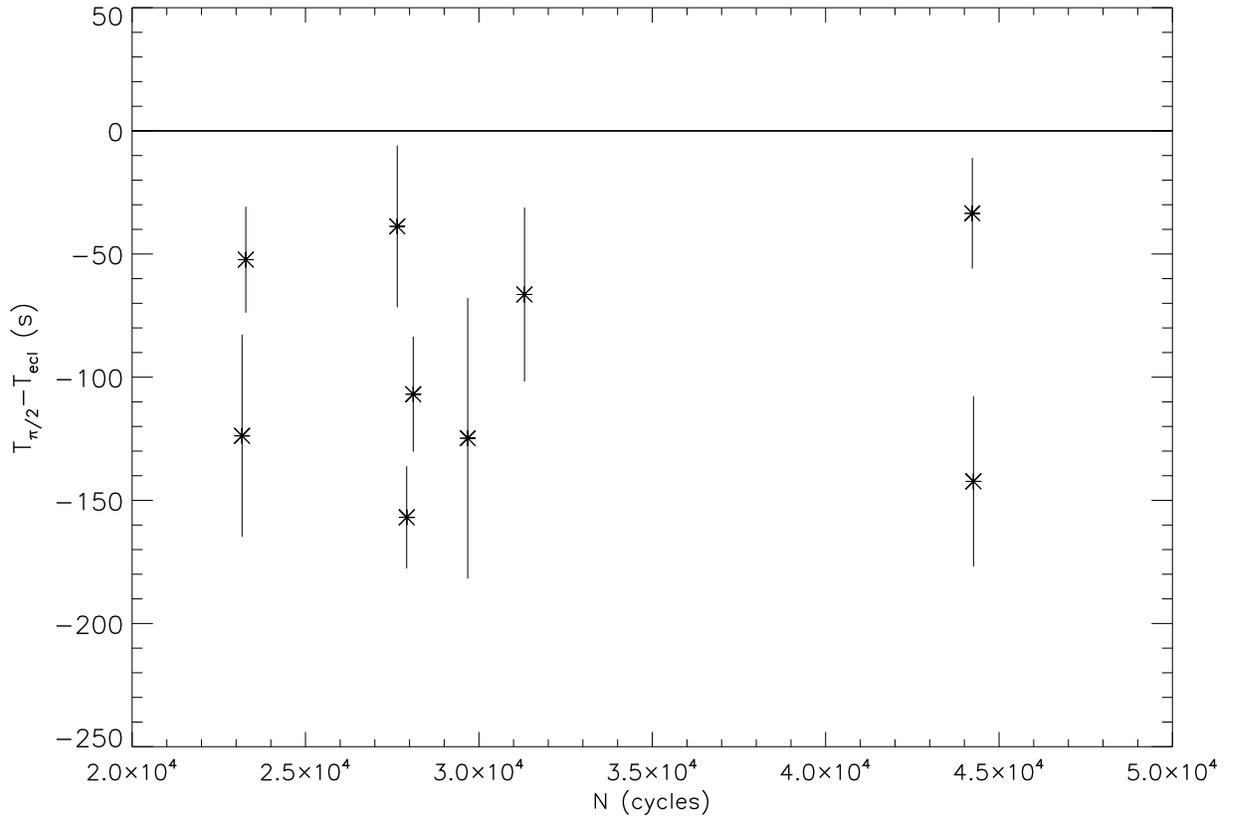}
\caption{Time differences between the measured $T_{\pi /2}$ values and expected X-ray eclipse times evaluated using the cubic ephemeris. The measured $T_{\pi /2}$ values are significantly earlier than the expected X-ray eclipse time by about 90 s. \label{t-cub}}
\end{figure}

However, significant difference between the expected X-ray eclipse times and $T_{\pi /2}$ can be seen in Figure~\ref{to-c}. Figure~\ref{t-cub} shows the time differences between the measured $T_{\pi /2}$ values and the eclipse times predicted by the cubic ephemeris (Eq~\ref{e_eph3}). The measured $T_{\pi /2}$ values are significantly and systematically earlier than the expected X-ray eclipse time by $90 \pm 11$ s (weight average). To further confirm this difference, we adopted an analysis method analogous to that used in \citet{iar11} for evaluating the time difference between the eclipse times measured from the X-ray and optical/UV bands. That is, we fixed the linear and quadratic terms as the quadratic ephemeris measured by the eclipses (Eq~\ref{e_eph2}) to fit both eclipse times and the measured the $T_{\pi /2}$ values. We found the phase zero epoch from eclipses is 45614.80949(12)MJD/TDB whereas 45614.80853(18)MJD/TDB form $T_{\pi /2}$, giving rise to a difference of $82 \pm 19$ s. We also did the same analysis but used the cubic ephemeris from eclipses (Eq~\ref{e_eph3}) and obtained that the phase zero epoch is 45614.80964(14)MJD/TDB form eclipses and 45614.80860(19)MJD/TDB form $T_{\pi /2}$, yielding a difference of $90 \pm 20$ s. The above results indicate that $T_{\pi /2}$ is significant earlier than X-ray eclipse times by $\sim 90$ s ($\sim 0.0045$ cycle). More discussions about this phenomenon are presented in section~\ref{dis_dt}.

\begin{figure}
\plotone{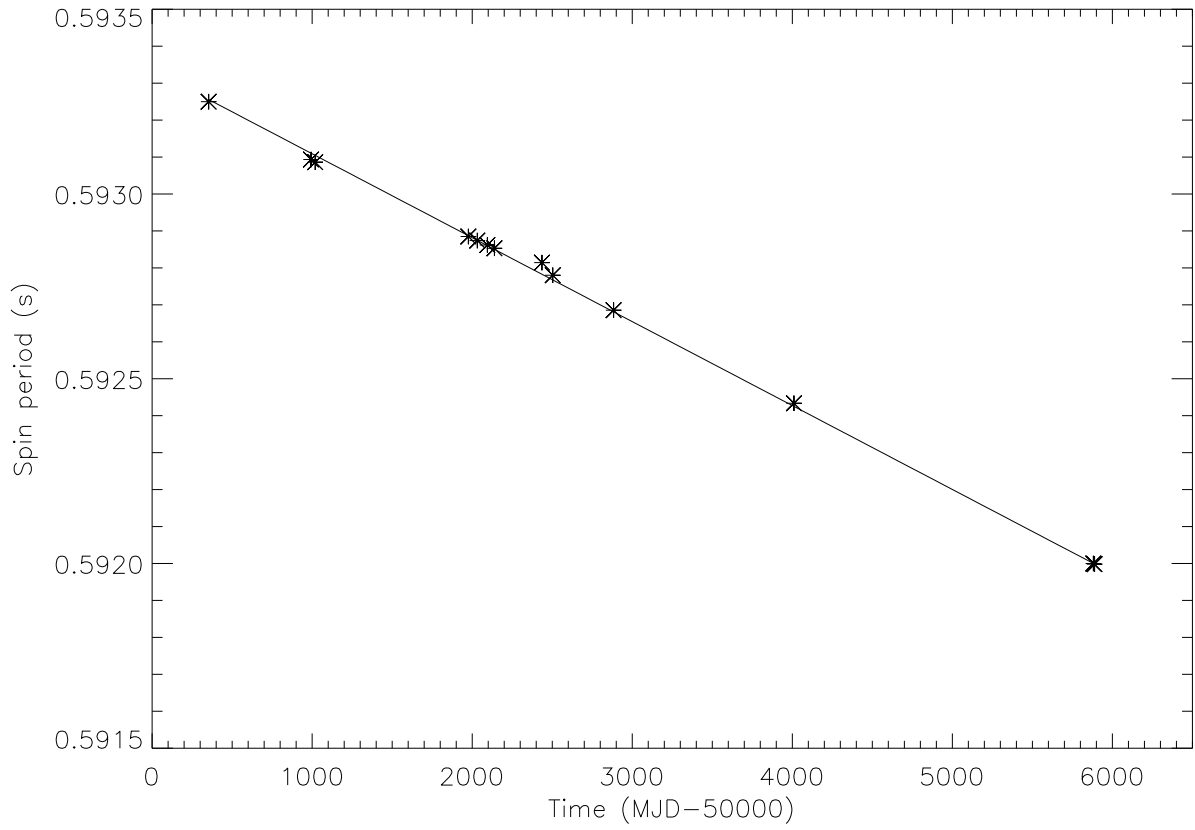}
\caption{Detected spin periods. The solid line is the best linear function to fit the evolution of the spin period. \label{spin}}
\end{figure}

The measured spin periods listed in Table~\ref{para} allow us to trace evolution of the neutron star spin period. To increase the statistics, the data sets that excluded in the evaluation of orbital and spin parameters because of insufficient number of data segments (i.e. less than 7 segments per data set) were retrieved to evaluate the spin parameters. Unfortunately, there was only one retrieved data set observed on June 7, 2002 with a sufficient number of data segments (larger than 2) to fit the spin parameters. We applied the same analysis method as we did for other data sets but with fixed orbital parameters for it. The orbital period was derived from the cubic ephemeris of X-ray eclipse times (Eq~\ref{e_eph3}), the $T_{\pi /2}$ value was estimated using the quadratic ephemeris (Eq~\ref{t_eph2}). We obtained a spin period of 0.5928144(26) s with phase zero epoch of $T_0=52435.67791044(15)$ MJD/TDB for this data set. Combined with the spin periods listed in Table~\ref{para} and the previous results from observations of {\it RXTE} in 1996 \citep{jon01}, {\it XMM-Newton} in 2001 \citep{iar15} and {\it Suzaku} in 2006 \citep{sas14}, we found the neutron star is spun-up with a rate of $\dot{P}_s=(-2.6288 \pm 0.0095) \times 10^{-12}$ s s$^{-1}$ (Figure~\ref{spin}), which is consistent with the value obtained by the Doppler effect analysis and close to those proposed by \citet{jon01}, \citet{jai10}, \citet{sas14}  and \citet{iar15}. The evolution of spin period can be described by

\begin{equation}\label{e_spin}
P_s(t)=0.5933359(41)s-2.6288(95) \times 10^{-12} \times (t-MJD50000) \times 86400, 
\end{equation}

\noindent where the time $t$ is in unit of MJD. All the errors have been scaled by a factor of $\sqrt{\chi^2_\nu}$ for conservatively estimating the uncertainties.

\section{Discussion} \label{dis}

We updated spin and orbital parameters for partial eclipsing LMXB X 1822-371 using X-ray eclipse times and pulsar timing. Combined with the X-ray eclipse times detected from {\it RXTE} 2011 observations and previously reported values with a time span of 34 years, we found that the evaluated orbital period derivatives decrease as the time span of the data increases, and the eclipse times is better described by a cubic ephemeris with a second order orbital period derivative of $\ddot{P}_{orb}=(-1.05 \pm 0.59) \times 10^{-19}$ s$^{-1}$. Pulsar timing gives an alternative way to measure the orbital and spin parameters of this system. The evolution of $T_{\pi / 2}$ values detected from 1998 to 2011 gives an orbital period derivative of  $(1.72 \pm 0.45)\times 10^{-10}$ s s$^{-1}$, consistent with that obtained from the quadratic ephemeris of eclipse times but they are significantly earlier than the expected X-ray eclipse times by $\sim 90$ s. Finally, we updated the spin-up rate to $\dot{P}_s=(-2.6288 \pm 0.0095) \times 10^{-12}$ s s$^{-1}$.

\subsection{Orbital Period Derivative} \label{dis_opd}
Although we found the orbital period derivative of X 1822-371 is decaying with a rate of $\ddot{P}_{orb}=(-1.05 \pm 0.59) \times 10^{-19}$ s$^{-1}$, the detected value from the quadratic model for the 34 years of time span shows that the orbital period derivative of $\dot{P}_{orb}=(1.464 \pm 0.041) \times 10^{-10}$ s s$^{-1}$ is still too large to be explained by the conventional orbital angular moment loss mechanisms caused by gravitational radiation and magnetic braking as proposed by \citet{bur10}. Such a large orbital period change rate is very likely driven by a significant amount of transferred mass lost from the binary system even if the accretion rate is close to the Eddington limit \citep{bur10,bay10}. A detected orbital period derivative significantly exceeding the theoretical prediction has been observed in some LMXBs. In addition to X 1822-371, a large orbital period change rate for a part of the LMXBs is also probably cause by mass outflow. For the ultra-compact LMXB X 1916-053, \citet{hu08} reported an orbital derivative of $\dot P_{orb} =(1.54 \pm 0.32) \times 10^{-11}$ s s$^{-1}$, about 200 times larger than that induced from garvitational radiation. \citet{hu08} estimated that  about 60\%-90\% of the mass lost from the companion is ejected from the binary system. It is probably caused by irradiation of the companion and accretion disk \citep{tav91}. Moreover, for the accreting millisecond X-ray pulsar SAX J1808.4-3658, \citet{dis08} found an unexpected large orbital period derivative of $\dot P_{orb} =(3.40 \pm 0.18) \times 10^{-12}$ s s$^{-1}$, a factor of 10 larger than that driven by gravitational radiation through a conservative binary mass transfer. \citet{dis08} proposed a mechanism similar to that of a black widow pulsar in that the mass outflow, even in a quiescence state, is induced by the pulsar wind. 

The radiation-driven mass transfer proposed by \citet{tav91} may explain the high orbital period derivative of  X 1822-371. Although the observed X-ray luminosity is only $\sim 10^{36}$ erg s$^{-1}$, the intrinsic X-ray luminosity may be as high as $10^{37}-10^{38}$ erg s$^{-1}$ for this ADC source and hence the expected mass loss from companion can be in the range of $\sim 10^{-8}-10^{-7}$ $M_{\sun}$ yr$^{-1}$ \citep{tav91}. If orbital angular momentum loss is caused by gravitational radiation, magnetic braking and mass outflow, then Eq(3) in \citet{dis08} can be rewritten as

\begin{equation}\label{dis08_3}
{\dot P_{orb} \over P_{orb}} =3\biggl\{{\biggl({\dot J_{orb} \over J_{orb}}\biggr)_{GR}}+{\biggl({\dot J_{orb} \over J_{orb}}\biggr)_{MB}}-{\dot M_2 \over M_2} \biggl[1-\beta q-\biggl(1-\beta \biggr)\biggl({{\alpha +q/3} \over {1+q} }\biggr) \biggr]\biggr\},
\end{equation}

\noindent where $\beta$ is the ratio of mass accreting onto the neutron star, $\dot M_1=-\beta M_2$, $\alpha$ is the fraction of the specific angular momentum loss from the companion star \citep[see][]{dis08}, mass ratio $q=M_2/M_1$, and $(\dot J_{orb} / J_{orb})_{GR}$ and $(\dot J_{orb} / J_{orb})_{MB}$ are orbital angular losses driven by gravitational radiation and magnetic braking, respectively. If the intrinsic X-ray luminosity can be written as $L_x=GM_1 \dot M_1 /R_1$ where $R_1$ is the radius of the neutron star, using the definitions of $\beta$ and mass ratio, the Eq~\ref{dis08_3} can be written as

\begin{equation}\label{dis08_3+1}
{\dot P_{orb} \over P_{orb}} =3\biggl\{{\biggl({\dot J_{orb} \over J_{orb}}\biggr)_{GR}}+{\biggl({\dot J_{orb} \over J_{orb}}\biggr)_{MB}}+ {{L_x R_1}\over {GM_1^2 q \beta} }   \biggl[1-\beta q-\biggl(1-\beta \biggr)\biggl({{\alpha +q/3} \over {1+q} }\biggr) \biggr]\biggr\}
\end{equation}

\begin{figure}
\plotone{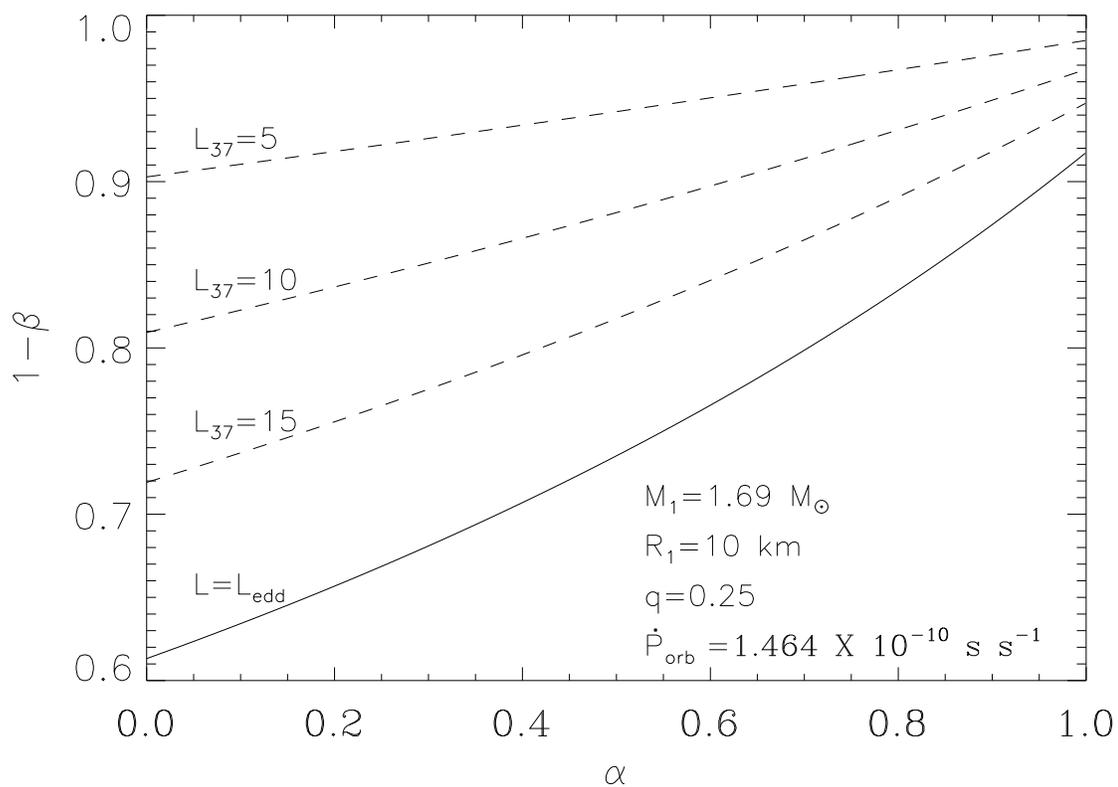}
\caption{The mass outflow ratio ($1-\beta$) as a function of fraction of specific angular momentum loss from from the companion star ($\alpha$) for a given X-ray intricsic luminosities. $L_{37}$ is the intricsic X-ray luminosity in unit of $10^{37}$ erg s$^{-1}$. \label{betaalpha}}
\end{figure}

The orbital angular momentum loss driven by gravitational radiation is given by  $(\dot J_{orb} / J_{orb})_{GR}=-32G^3M_1M_2(M_1+M_2)/(5c^5a^4)$, where $c$ is speed of light and $a$ is the binary separation evaluated by Kepler's third law $a=[G(M_1+M_2)/4\pi^2]^{1/3}P_{orb}^{2/3}$. Taking the neutron star mass $M_1=1.69M_{\sun}$ as suggested by \citet{iar15}, and mass ratio $q=0.25$ based on $0.24 \leq q \leq 0.27$ proposed by \citet{mun05}, we found $(\dot J_{orb} / J_{orb})_{GR}=-4.40 \times 10^{-11}$ yr$^{-1}$. Moreover, the orbital angular momentum loss driven by magnetic braking can be estimated estimated by Eq(4) in \citet{ver81}, $\dot J=-0.5 \times 10^{-28} f^{-2}k^2M_2R_2^4\Omega_{orb}$ where $\Omega_{orb}$ is the angular velocity of orbital motion and $R_2$ is the radius of the companion, approximately equal to the radius of the Roche lobe, i.e., $R_2 \approx R_{L}=1/3^{4/3}[1/(1+q)]^{1/3}a$ \citep{pac71}. Taking $k^2=0.1$ as suggested by \citet{ver81} and $f=0.73$ according to \citet{sku72}, we obtained  $(\dot J_{orb} / J_{orb})_{MB}=-2.62 \times 10^{-9}$ yr$^{-1}$. Combining the observed orbital period derivative $\dot P_{orb}/P_{orb}=2.30 \times 10^{-7}$ yr$^{-1}$ from the quadratic ephemeris and $R_1=10^6$ cm, then substituting these numerical values into Eq~\ref{dis08_3+1}, we obtained the mass outflow ratio (i.e., $1-\beta$) as a function of $\alpha$ for a given intrinsic luminosities $L_x$, as shown in Figure~\ref{betaalpha}. Therefore, at least $\sim 60\%$ of the mass lost form the companion has to be ejected from the binary system. If the accretion rate is close to the Eddington limit, the mass loss rate from companion is $\dot M_2=3.86 \times 10^{-8}$ $M_{\sun}$ yr$^{-1}$ for $\alpha=0$ ($\beta=0.39$) and   $\dot M_2=1.81 \times 10^{-7}$ $M_{\sun}$ yr$^{-1}$ for $\alpha=1$ ($\beta=0.083$), which agree with the range of $10^{-8}-10^{-7}$ $M_{\sun}$ yr$^{-1}$ for the radiation-driven mass transfer mechanism \citep{tav91}.

On the other hand, \citet{hel90} proposed that the large, positive orbital period derivative may be caused by short-term effects departing from the long-term evolutionary trend. One of the possible short-term effects is that magnetic, solar-type cycles of the companion affect its mass distribution, and hence the orbital period changes are a result of variations in the quadrupole moment \citep{hel90}. This model has been applied to explain the orbital period glitches observed in EXO 0748-676, a total eclipsing LMXB consisting of a neutron star and a $0.45 M_{\sun}$ low mass main-sequence companion \citep{par86}, with an orbital period of 3.82 hr.  \citet{wol09} analyzed 433 full X-ray eclipses and found that the orbital period has been experiencing abrupt changes with a time scale of several milliseconds three times during a data time span of 23 years. Combined with a possible detection of magnetic loop structures of the companion \citep{wol07}, \citet{wol09} proposed that the magnetic activity cycles, as suggested by \citet{her97} are likely to be responsible for the orbital glitches. Because the mid of eclipse times of this total eclipse system can be determined with a high accuracy, evolution of the orbital period can also be precisely traced. In contrast, the uncertainty of the phase of the fiducial point for a partial eclipsing system such as X 1822-371 is too large ($\sim 10^{-3}$ cycle) compared with that measured by total eclipse \citep[$\sim 10^{-5}$ cycle, see Table 1 in][]{wol09}, and such a small amount of orbital period change is hard to be directly observed in X 1822-37 \citep{wol02}. However, \citet{wol09} reported a net orbital period changed of +9.16 ms during the 23 years time span (from MJD46110 to 54647). The average orbital derivative ($\Delta P_{orb} / \Delta t$) is $1.24 \times 10^{-11}$ s s$^{-1}$, an order of magnitude smaller than that of X 1822-371 obtained from the quadratic ephemeris ($1.464 \times 10^{-10}$ s s$^{-1}$). That is, if the large, positive orbital period derivative of X 1822-371 was mainly caused by magnetic activity cycles of the companion, the activity would be either 10 times more frequently or 10 times stronger than those in the companion of EXO 0748-676. We therefore conclude that this model is insufficient to explain the orbital derivative of X 1822-371.

\begin{figure}
\plotone{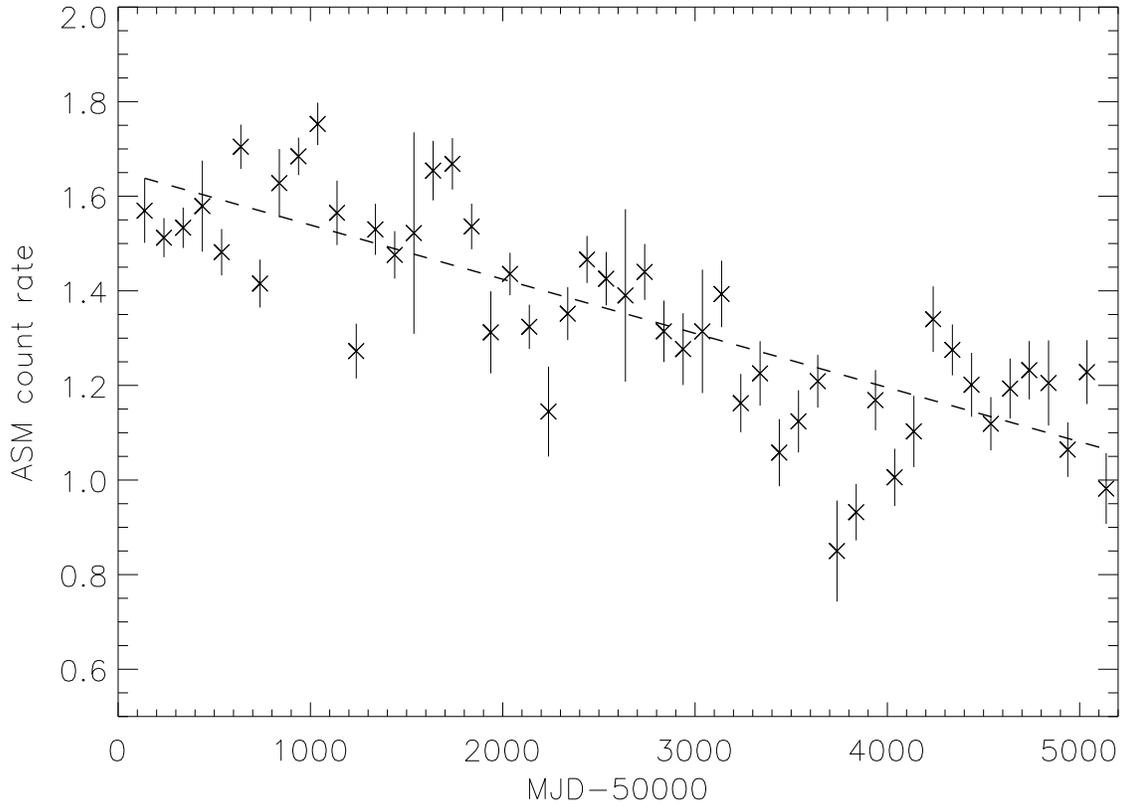}
\caption{The 2-12 keV X-ray light curve collected by the All Sky Monitor onbroad the {\it RXTE} from 1996 to 2009 with a data bin size of 100 d. A clear declining trend can be observed. The dashed line shows the linear declining trend of the light curve with a mean decline rate of -0.0418$\pm$0.0044 cts/s per year. \label{asmlc}}
\end{figure}

However, we found that the cubic ephemeris probably better describes evolution of the eclipse times of X 1822-371 than the quadratic ephemeris. If the detection is true, form the detected second order orbital period derivative $\ddot{P}_{orb}=-1.05 \times 10^{-19}$ s$^{-1}$, the change in orbital period derivative during the $\sim$34 years data time span is $\Delta (\dot P_{orb}/P_{orb})=-1.78 \times 10^{-7}$ yr$^{-1}$, decreasing by $\sim 60\%$ compared with the orbital period derivative at the beginning of the data (MJD43413.0272) evaluated from the cubic ephemeris. Such a large orbital derivative change is not likely driven by gravitational radiation and magnetic braking because their contributions to the orbital period derivative are only $3[(\dot J_{orb} / J_{orb})_{GR}+(\dot J_{orb} / J_{orb})_{MB}]=-7.99 \times 10^{-9}$ yr$^{-1}$ (see Eq~\ref{dis08_3}). Thus, the orbital period derivative change can be only driven by the mass loss/outflow variation (the third term on the right-hand side of Eq~\ref{dis08_3} or ~\ref{dis08_3+1}). If the mass loss of the companion is mainly driven by radiation, a lower intrinsic X-ray luminosity may reduce the mass loss/outflow rate, even though the relation between the outflow ratio (or $\beta$ value) and intrinsic X-ray luminosity is complicated \citep{tav91}. A supporting evidence for decreasing of intrinsic X-ray luminosity can be observed in the X-ray light curve of X 1822-371 collected by the All Sky Monitor onboard the {\it RXTE}. It shows that the detected count rate in the 2-12 keV band was declining with time (see Figure~\ref{asmlc}) although the time span for the light curve is only $\sim$14 years instead of $\sim$34 years for the measured eclipse times and the energy range for the light curve is only 2-12 keV. This implies that the accretion rate is likely decreasing, resulting in a negative second order orbital period derivative. However, to further verify if the accretion rate was declining during the $\sim$34 years, a light curve for the bolometric flux variation is required.

\subsection{Deviation between $T_{\pi / 2}$ and X-ray Eclipse Time} \label{dis_dt} 
The evolution of $T_{\pi / 2}$ derived from pulsar timing in section~\ref{pa} gives an independent measurement of the orbital period derivative of X 1822-371, which is consistent with that derived from the quadratic ephemeris of X-ray eclipse times except for a significant $\sim 90$ s deviation between them. Similar deviation can also be seen between the eclipse times in the X-ray and optical/UV bands. \citet{whi81} first reported that X-ray eclipse time lags the optical minimum by about 0.04 cycle ($\sim$13 min) but this value was corrected to $3.0 \pm 3.4$ min \citep{hel89} and then further refined to $180 \pm 50$ s \citep{hel90}, $100 \pm 65$ s \citep{bay10} and $127 \pm 52$ s \citep{iar11}, although the delay may not be a constant \citep{iar11}. This deviation is very likely caused by different emission regions of the X-ray and optical/UB photons. The X-rays are emitted from the ADC around the neutron star, whereas the optical/UV photons are mainly from the asymmetry accretion disk \citep{hel89}.

We discovered the X-ray eclipse time lags $T_{\pi / 2}$  by $90 \pm 11$ s. This deviation is probably caused by asymmetric X-ray emissions to the observers. The observed X-rays are scattered from the ADC and partially obscured by the outer rim of the accretion disk \citep{whi82,hel89}. The asymmetry may be caused by asymmetric absorption of the disk rim or X-ray emissions from the ADC; hence the centroid of light is offset from the line connecting the centers of mass of the primary and secondary stars in the binary system, resulting in the deviation between X-ray eclipse time and $T_{\pi / 2}$ for a circular orbit. \citet{jon03} measured the radial velocity for the companion of X 1822-371 using the \ion{He}{1} absorption lines at 4026.357 and 5879.966 $\rm{\AA}$ and found the minimum of radial velocity occurs earlier than the expected one evaluated using the ephemeris from pulsar timing by $0.08 \pm 0.01$ cycle ($\sim$1600 s), likely a result of asymmetric heating of the companion star. However, \citet{cas03} detected the radial velocity using the Doppler imaging of the fluorescent \ion{N}{3} $\lambda$4640 emission line and found that the radial velocity curve agrees well with that anticipated from pulsar timing with no evidence for asymmetric irradiation of the companion. The deviation reported by \citet{jon03} is caused by the \ion{He}{1} absorber located at the leading side of the companion's Roche lobe or over the gas stream \citep{cas03}. Our discovery of the deviation between the eclipse time and $T_{\pi / 2}$ suggests that the X-ray emissions are asymmetric but the degree of asymmetry is much smaller than that proposed by \citet{jon03}. This small deviation is probably below the sensitivity of the measurment method used in \citet{cas03}.

On the other hand, it is also possible that the deviation between the eclipse time and $T_{\pi / 2}$ is caused (or partly caused) by a small eccentricity of the binary orbit. From Eq (3) in Appendix II of \citet{van84}, the relation between $T_{\pi / 2}$ and the time of superior conjunction, $T_{conj}$ for a small orbital eccentricity $e$, can be written as

\begin{equation}\label{van84_3}
T_{\pi / 2}=T_{conj}+{{eP_{orb}} \over {\pi}} \cos \omega,
\end{equation}

\noindent where $\omega$ is the periastron angle. If we assume the X-ray emissions from the primary is symmetric, using Eq (4) in in Appendix II of \citet{van84}, the relation between eclipse time $T_{ecl}$ and  $T_{conj}$ is

\begin{equation}\label{van84_4}
T_{ecl}=T_{conj}-{{eP_{orb}} \over {\pi}} \cos \omega {{(\sin i - \beta)(1-\beta \sin i)} \over {\beta \sin ^2 i}},
\end{equation}

\noindent where $\beta \equiv [1-(R/a)^2(1-e^2)^{-1}]^{1/2}$, $i$ is inclination angle, $R$ is the radius of companion and $a$ is the binary separation. Combining Eq~\ref{van84_3} and  ~\ref{van84_4}, the relation between  $T_{\pi / 2}$ and $T_{ecl}$ can be written as

\begin{equation}\label{van84_3+4}
T_{ecl}-T_{\pi / 2}={{eP_{orb}} \over {\pi}} \cos \omega {\Bigl [ }1-{{(\sin i - \beta)(1-\beta \sin i)} \over {\beta \sin ^2 i}} {\Bigr ]},
\end{equation}  

For X 1822-371, we adopted $i=82{\degr}.5$ \citep{hei01}, $R/a=2/3^{4/3}[q/(1+q)]$ \citep{pac71}, the mass ratio $q=0.25$ (see section~\ref{dis_opd}) and $\beta \approx [1-(R/a)^2]^{1/2}$ (to first order in $e$). The value of $(\sin i -\beta)(1-\beta \sin i) /(\beta \sin^2 i)$ in Eq~\ref{van84_3+4} is only $1.7 \times 10^{-4}$ and thus we find

\begin{equation}\label{van84_3+4app}
e \cos \omega \approx {{\pi (T_{ecl}-T_{\pi / 2})} \over {P_{orb}}}=0.014,
\end{equation} 

\noindent which is smaller than the upper limit of the eccentricity of X 1822-371 \citep[0.031,][]{jon01}. We therefore cannot exclude the possibility that the deviation between the eclipse time and $T_{\pi / 2}$ is caused by the small eccentricity of the binary orbit.

\subsection{Constraints on Mass and Radius of the Neutron Star in X 1822-371} \label{dis_upl}

\citet{gho79} have discussed the relation between the spin-up time scale $T_s$  and magnetic dipole moment $\mu$. For a specific source with a spin period $P_s$, X-ray luminosity $L_x$ and certain neutron star model, they pointed out that the expected $T_s$ is only a function of $\mu$, and that $T_s(\mu)$ decreases as $\mu$ increases and then rises after passing a minimum value \citep[see, for example, Fig. 12 in][]{gho79}. Compared with the observed time scale $\bar{T}_s$, there are two solutions, i.e., slow rotator solution with a smaller fastness parameter $\omega_s \equiv \Omega_s / \Omega_k (r_{in})$, where $\Omega_s$ is the spin angular frequency and $\Omega_k (r_{in})$ is the Keplerian angular frequency of the inner radius of the accretion disk, and a fast rotator solution with a larger $\omega_s$. 

Because $T_s(\mu)$ has a minimum value, it implies that the expected spin-up rate $(-\dot P_s(\mu))$ has a maximum value $(-\dot P_s)_{max}$ for a specific source. The observed spin-up rate  $(-\dot P_s)_{obs}$ must be smaller than this maximum value, that is, $(-\dot P_s)_{obs} \le (-\dot P_s)_{max}$. This allows us to constrain the mass and radius of the neutron star in X 1822-371 for a given luminosity and  moment of inertia as functions of mass and radius of neutron star, independent of the magnetic moment of the neutron star.  Eq.(15) in \citet{gho79} is be rewritten as 

\begin{equation}\label{gho79_15}
-\dot P_s=1.6 \times 10^{-12} \mu_{30}^{2/7} n(\omega_s)S_1(P_s L_{37}^{3/7})^2 \ s \ s^{-1},
\end{equation}

\noindent where $S_1=R_6^{6/7}(M_1/M_{\sun})I_{45}^{-1}$, $\mu_{30}$ is the magnetic moment in units of G cm$^3$, $P_s$ is the spin period of the neutron star in units of second, $L_{37}$ is the luminosity in units of $10^{37}$ erg/s, $R_6$ is the radius of the neutron star in units of $10^6$ cm, $M_1$ is the mass of neutron star, and $I_{45}$ is the moment of inertia of the neutron star in units of $10^{45}$ g cm$^2$. We applied the approximate value of dimensionless accretion torque from Eq. (10) in \citet{gho79}, $n(\omega _s) \approx 1.39{1-\omega_s [4.03(1-\omega_s)^{0.173}-0.878]}(1-\omega_s)^{-1}$, which is accurate to 5\% for $0 \le \omega_s \le 0.9$ and the fastness parameter from  Eq. (16) and (18) in \citet{gho79}, $\omega_s \approx 1.35 \mu_{30}^{6/7}R_6^{-3/7}(M_1/M_{\sun})^{-2/7}(P_s L_{37}^{-2/7})^{-1}$. The moment of inertia was adopted from the Eq. (12) in \citet{lat05}, which is good for $M_1 \ge 1 M_{\sun}$. 

\begin{figure}
\plotone{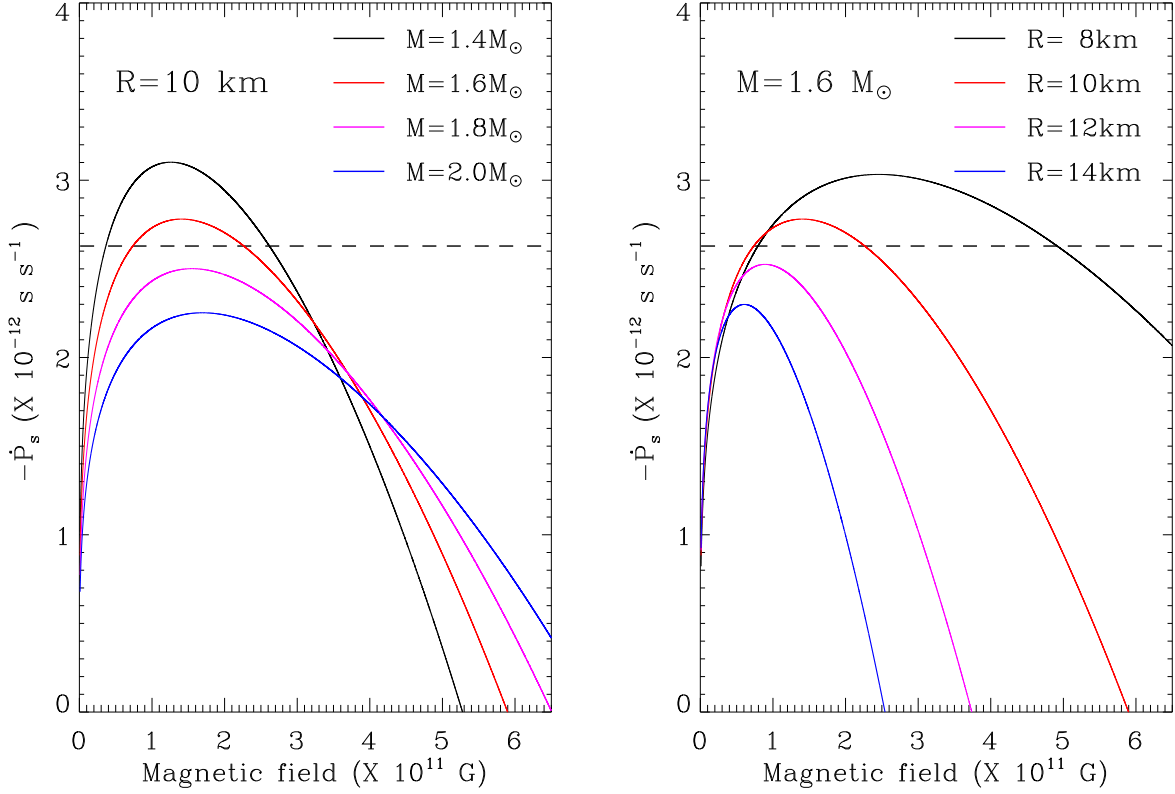}
\caption{Spin-up rate as a function of neutron star magnetic field for various masses with the radius fixed (left) and various radii with the mass fixed (right). The dashed lines are the observed spin-up rate ($-2.6288 \times 10^{-12}$ s s$^{-1}$) derived from this work. These two plots indicate that the upper limit of the neutron star mass of X 1822-371 is between 1.6 $M_{\sun}$  and 1.8 $M_{\sun}$ for a radius of 10 km and the upper limit of the neutron star radius is between 10 km and 12 km for a mass of 1.6 $M_{\sun}$. \label{spinup3}}
\end{figure}

\begin{figure}
\plotone{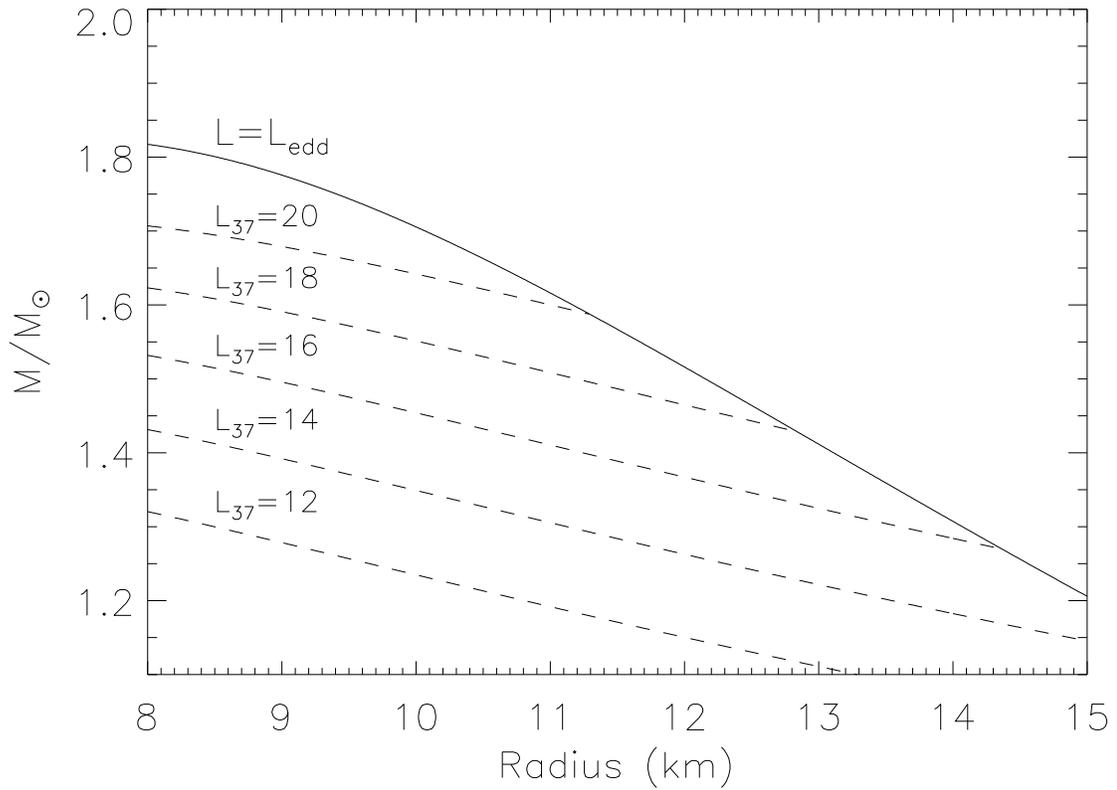}
\caption{Upper limit curves of mass and radius for the neutron star in X 1822-371. The solid line shows the upper limit for the case that the intrinsic luminosity equals to the Eddington luminosity. The dashed lines are the upper limits for various fixed intrinsic luminosities. \label{mrupperlim}}
\end{figure}

Figure~\ref{spinup3} shows the spin-up rate of the neutron star in X 1822-371 as a function of neutron star magnetic field for various masses with the radius fixed as well as for various radii with the mass fixed, under the assumption that the intrinsic luminosity is the Eddington luminosity. Compared with the observed spin-up rate, it can give the upper limits of the mass and radius of the neutron star. Figure~\ref{mrupperlim} shows the upper limit curves of the mass and radius of the neutron star in X 1822-371 for that the intrinsic luminosity equals to its Eddington luminosity and various fixed luminosities.  This figure also implies that the intrinsic luminosity of  X 1822-371 is likely $\gtrsim 10^{38}$ erg s$^{-1}$ for $\dot{P}_s=-2.6288  \times 10^{-12}$ s s$^{-1}$; otherwise it would give an unreasonable upper limit for the mass and radius of the neutron star. This is consistent with that the intrinsic luminosity of X 1822-371 could be close to its Eddington luminosity suggested by \citet{whi82,bur10} and \citet{bay10} based on its unusual large orbital period derivative.

\acknowledgments

The {\it RXTE} data for this research were obtained from the High Energy Astrophysics Science Archive Research Center (HEASARC) online service, provided by the NASA/Goddard Space Flight Center. This work was supported by the Ministry of Science and Technology of Taiwan through the grants NSC 102-2112-M-008-020-MY3 and MOST 105-2112-M-008-012-. C.-P. H. especially acknowledges the support provided by an ECS grant of Hong Kong Government under HKU 70971.

\clearpage

\clearpage

\clearpage

\clearpage

\clearpage

\clearpage

\clearpage

\clearpage

%\clearpage

%\begin{figure}
%\plotone{spinup1.eps}
%\caption{Example of the relation between the spin-up rate and neutron star magn;etic field for $M=1.64 M_{\sun}$, $R=10$ km, and with the Eddington accretion rate. It was made to match the observed neutron star spin-up rate for X 1822-371 in this work and its possible magnetic field $8.8 \times 10^{10}$ G proposed by \citet{iar15}. Because the observed spin-up rate in this work is slightly larger than that reported by \citet{iar15}, the neutron mass has to be reduced from $1.69 M_{\sun}$ to $1.64 M_{\sun}$. There are a maximum spin-up rate and two solutions for the observed spin-up rate where the possible magnetic field  proposed by \citet{iar15} is the slow rotator solution.\label{spinup1}}
%\end{figure}

\clearpage

\clearpage
\begin{table}
\begin{center}
\caption{X-ray eclipse times detected from RXTE 2011 observations\label{etime}}
\begin{tabular}{ccc}
\\
\tableline\tableline
Eclipse time & Error (d) & Cycle count\\
(MJD/TDB) & & \\
\tableline
55881.01759 & 0.00029 & 44230 \\
55881.71527 & 0.00034 & 44233 \\
55884.73103 & 0.00049 & 44246 \\
55888.44553 & 0.00035 & 44262 \\
\tableline
\end{tabular}
\end{center}
\end{table}

\clearpage
\begin{table}
\begin{center}
\caption{Best-fitted orbital parameters from the X-ray eclipse times of X 1822-371  \label{efit}}
\begin{tabular}{lll}
\\
\tableline\tableline
Parameter\tablenotemark{a} & Quadratic & Cubic\\
          & Model\tablenotemark{b}     & Model\tablenotemark{c}\\
\tableline
a (d)            & $(7.3 \pm17.2) \times 10^{-5}$   & $(2.1 \pm 1.8) \times 10^{-4}$ \\
b (d)            & $(-1.48 \pm 0.21) \times 10^{-7}$& $(-2.37 \pm 0.54) \times 10^{-8}$\\
c (d)            & $(1.700 \pm 0.048) \times 10^{-11}$& $(2.25 \pm 0.31) \times 10^{-11}$\\
d (d)            & -----                          & $(-8.2 \pm 4.6) \times 10^{-17}$ \\
$T_{0,orb}$ (MJD/TDB) & 45614.80949(17)            &  45614.80964(18) \\
$P_{orb}$ (d)       & 0.232108983(91)            &  0.232108780(54)\\
$\dot{P}_{orb}$ (d) & $(1.464 \pm 0.041) \times 10^{-10}$ & $ (1.94 \pm 0.27) \times 10^{-10}$\\
$\ddot{P}_{orb}$ (d) & -----                      & $(-9.1\pm 5.1) \times 10^{-15}$\\ 
$\chi^2/(d.o.f.$) &53.72/30                       & 48.42/29                 \\ 
\tableline
\tablenotetext{a}{All the errors of parameters have been scaled by a factor of $\sqrt{\chi^{2}_{\nu}}$.}
\tablenotetext{b}{$\Delta t=a+bN+cN^2$, where $\Delta t$ is the time delay of the observed eclipse times in comparison to the linear ephemeris proposed by \citet{hel94} and $N$ is the cycle count.}
\tablenotetext{c}{$\Delta t=a+bN+cN^2+dN^3$.}
\end{tabular}
\end{center}
\end{table}

\clearpage
\begin{table}
\begin{center}
\caption{Detected orbital period derivative of X 1822-371 from X-ray eclipse times \label{ep_dot}}
\begin{tabular}{ccccl}
\\
\tableline\tableline
Detected Orbital Period  &  Start Time & Stop Time & Time Span & Reference\\
Derivative ($\times 10^{-10}$ s s$^{-1}$)& (MJD/TDB) & (MJD/TDB) & (d) &  \\
\tableline
$2.19 \pm 0.58$ &43413.02720 & 47759.72900 & 4346.70180 & \citet{hel90} \\
$2.04 \pm 0.48$ &43413.02720 & 48692.34396 & 5279.31676  & \citet{hel94} \\
$1.78 \pm 0.20$ &43413.02720 & 50701.01870 & 7287.99150  & \citet{par00} \\
$1.499 \pm 0.071$ &43413.02720 & 54607.19592 & 11194.16872  & \citet{bur10} \\
$1.514 \pm 0.080$ &43413.02720 & 54609.74890 & 11196.72170  & \citet{iar11} \\
$1.464 \pm 0.041$ &43413.02720 & 55888.44507 & 12475.41787  & This work\\

\tableline
\end{tabular}
\end{center}
\end{table}

\clearpage
\begin{deluxetable}{lcllllll}
\tabletypesize{\scriptsize}
\rotate
\tablecaption{Best-Fit orbital and spin parameters\label{para}}
\tablewidth{0pt}
\tablehead{
\colhead{Observation Time} &Observation ID& \colhead{$P_{orb}$\tablenotemark{a}} &\colhead{$a_x \sin i$} & \colhead{ $T_{\pi /2}$} &\colhead{$T_0$\tablenotemark{b}}&\colhead{$P_s$\tablenotemark{c}}& \colhead{$\dot \nu_s$\tablenotemark{d}}\\

       &  &\colhead{(s)}     & \colhead{(lt-s)}      & \colhead{(MJD/TDB)}   &\colhead{(MJD/TDB)}& \colhead{(s)} &\colhead{(Hz s$^{-1}$)}

}
\startdata

1998/06/28-29 & 30060 & 20054.27736530(1) & 0.988(10)  & 50992.77972(48)  & 50992.80499978(18) & 0.59309334(11) & -   \\

1998/07/24-25 & 30060 & 20054.27768510(1) & 0.9945(83) & 51018.31261(25) & 51018.90117848(13)  & 0.593086164(81) & -  \\

2001/05/02-03 & 50048 & 20054.28996210(1) & 1.011(17)  & 52031.70401(38) & 52031.75939201(13)   & 0.59287396(13)  & -  \\

2001/07/01-05 & 50048 & 20054.29070020(1) & 0.9957(64)   & 52094.83625(24) & 52094.700018432(49)  & 0.592861866(48) & $8.6(1.1) \times 10^{-12}$  \\

2001/08/17-20 & 60042 & 20054.29121130(1) & 1.0261(68) & 52138.70586(27) & 52138.74689145(16) & 0.592852949(33) & -  \\

2002/08/02-17 & 70037 & 20054.29540060(1) & 0.994(14)  & 52503.35027(66) & 52503.34734500(11) & 0.59278016(13) & $1.07(4) \times 10^{-11}$  \\

2003/08/31-09/14 & 70037 & 20054.29965470(1) & 0.9783(97) & 52883.31289(41) & 52883.28899623(25) & 0.59268544(15) & $8.70(72) \times 10^{-12}$  \\

2011/11/15-16 & 96344 & 20054.32923720(4) & 1.0092(86) & 55881.01757(26) & 55881.103830939(58) & 0.5900044(80) & -  \\

2011/11/23-30 & 96344 & 20054.32930370(4) & 1.010(11) & 55888.67595(40) & 55888.50670071(13) & 0.591998118(17) & -  \\
\enddata
\tablenotetext{a}{Orbital period, evaluated from the cubic ephemeris of X-ray eclipse times (Eq~\ref{e_eph3}) and kept as constants for perameter corrections}
\tablenotetext{b}{Phase zero epoch of pulsation}
\tablenotetext{c}{Spin period of neutron star}
\tablenotetext{d}{Spin frequency derivative, required for some data sets to obtain better fitting}
\end{deluxetable}

\end{document}